\documentclass[5p]{elsarticle}

\usepackage{ifpdf}
\ifpdf\usepackage{epstopdf}\fi
\usepackage{amsmath,amssymb}

\usepackage{lineno,hyperref}
\modulolinenumbers[5]
\usepackage{color}

\journal{Journal of \LaTeX\ Templates}

\biboptions{sort&compress}

\begin{document}

\begin{frontmatter}

\title{Modeling of liquid metal droplet deformation by laser impact}

\author{I.Yu. Vichev\corref{mycorrespondingauthor}}
\cortext[mycorrespondingauthor]{Corresponding author}
\ead{vichevilya@keldysh.ru}

\author{D.A. Kim}
\address{Keldysh Institute of Applied Mathematics RAS, Miusskaya sq.4, 125047 Moscow, Russian Federation}

\author{V.V. Medvedev}
\address{Institute of spectroscopy RAS, Fizicheskaya~str., 5, Troitsk, Moscow, 108840, Russia}

\begin{abstract}
The method of sequential simulation of liquid metal droplet deformation by a laser pulse is considered. The first stage is the laser impact on a droplet. It was simulated using RALEF-2D code, based on the radiative gas dynamic model. The next stage is target deformation from a droplet to a disk. This part of simulation was carried out using OpenFOAM code where surface tension forces are taken into account. Good agreement with experimental results was obtained.
\end{abstract}


\end{frontmatter}


\section{Introduction}
In modern lithographic facilities, the most important element is a source of EUV radiation with a wavelength of 13.5 nm $\pm$ 1\% based on tin plasma~\cite{book:666357, 0022-3727-44-25-253001}, obtained by the impact of a high-intensity laser pulse (of the 10$^{10}$ W/cm$^2$ order) on a distributed tin target~\cite{Koshelev2012}. At this stage in the development of EUV technology, the most promising is the use of a distributed target in the form of a thin liquid tin disk with a diameter of several hundred micrometers. To prepare such a target, a laser prepulse of lower power ($\sim$~10$^9$~W/cm$^2$) with a characteristic duration of the order of 10~--~100~ns and with a smooth intensity distribution in time and space is used. It impacts on liquid tin droplets with a diameter of several tens of micrometers, emitted from the droplet generator with a high frequency ($\sim$~50~kHz), and initiates their transformation into a disk~\cite{Brandt2014}.

Each act of generating EUV radiation can be divided into three stages. The first stage is the irradiation of a tin drop with a laser pulse. The drop begins to deform and sets in motion. In the second stage, the target moves with constant acquired speed by inertia and stretches into a disk. The third stage begins when the main, more powerful, the laser pulse is turned on, which converts the target disk into a hot plasma, which is the source of EUV radiation. In this case, the target burns out almost completely. This process is repeated at a certain frequency determined by the droplet generator and the laser system operation. The effectiveness of such a source directly depends on the quality of preparation of the distributed target. Understanding the physics of this process and the ability to make predictions about how laser pulse parameters affect a future target is very much in demand for improving settings. The development and improvement of physical models, methods, and program codes allow such predictions to be made is an extremely urgent task.

The study of phenomena occurring at the third stage is widely covered~\cite{Murakami2006, Basko2016} and they are not considered here. This work is devoted to numerical modeling of the interaction of a laser pulse with a liquid metal droplet (LMD) and its subsequent deformation into a disk-shaped target, i.e., the subject of the study is the first two stages of the act of generating EUV radiation. For this, a method was developed for the sequential use of two different software codes, RALEF-2D~\cite{Basko_developmentof, Basko2012} and OpenFOAM~\cite{Weller1998}. To simulate the processes occurring in the first stage, where the interaction of radiation with matter plays a significant role, the RALEF-2D code is used. At the second stage, the influence of surface tension forces on the LMD deformation becomes significant, and radiation transfer and thermal conductivity after turning off the laser stop to play a noticeable role, therefore, modeling continues in the open platform OpenFOAM for numerical simulation of continuum mechanics problems.

The authors developed a special algorithm for transferring the results of calculating velocity and density fields obtained using RALEF-2D at the end of the first stage of the process in OpenFOAM as initial data for modeling the second stage, in compliance with the laws of conservation of mass and momentum.

To verify this technique, we used the results of experiments conducted at the ARCNL center at the University of Amsterdam (Netherlands)~\cite{PhysRevApplied.6.014018}.

\section{Statement of the problem}
\label{section_Statement}

The paper presents the results of a series of experiments on the interaction of an Nd: YAG laser with a wavelength of 1064~nm and a tin drop with a diameter of 50~$\mu$m~\cite{PhysRevApplied.6.014018}. It was decided to verify the studied method for modeling the deformation of the target by comparing the simulation with two cases from the series of experiments indicated on Fig. 4 in \cite{PhysRevApplied.6.014018} as ``We = 20'' and ``We = 336''. Where $\text{We} = \rho R_{0} U^{2} / \sigma$ is Weber number, $\rho$~=~6.78~g/cm$^3$ and $R_0$~=~25~$\mu$m are the density and radius of the initial drop, respectively, $U$ is the target velocity acquired as a result of the pulse, and $\sigma$~=~538~mN/m is the surface tension of liquid tin. The simulation performed using RALEF-2D code for ``We = 20'' and ``We = 336'' cases with laser beam energy 0.63~mJ and {\color{red}X.X~mJ} correspondingly. The laser pulse intensity in the simulation had a Gaussian circular distribution with a spot diameter of 115~$\mu$m (FWHM) and a Gaussian distribution over time with a duration of 10~ns (FWHM).

\section{Stages of the simulation process}
\label{section_Stages}

This section provides a more detailed description of the processes leading to the deformation of LMD by impact of the laser pulse occurring at the first and second stages of EUV generation.

\subsection{Stage 1: the laser beam interaction with liquid metal drop}
\label{subSection_Stage1}

\begin{figure}[!htb]
    \centering
    \includegraphics[width=0.5\linewidth]{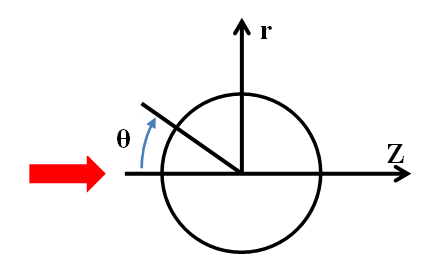}
    \caption{Scheme of the process under consideration.}
    \label{Scheme_process}
\end{figure}

\begin{figure}[!htb]
    \centering
    \includegraphics[width=0.4\linewidth]{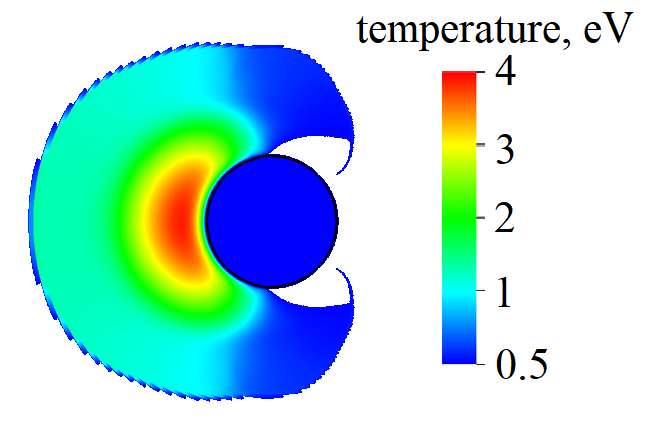}
    \caption{The temperature of plasma at time 20 ns.}
    \label{temperature_20ns}
\end{figure}

At the first stage, the interaction of laser with an LMD occurs (see Fig.~\ref{Scheme_process}). As a result of a laser action, the surface of the droplet heats up. Since the average power of the considered laser pulse is higher than the ablation threshold ($\sim$ 10$^8$ W/cm$^2$)~\cite{Basko2015}, the interaction process quickly (within a few nanoseconds) switches to the ablation mode. The temperature on the irradiated part of the surface of the drop rises above the boiling point of tin ($\sim$ 2800 K), and intense evaporation of the substance begins. The sharp boundary between liquid and vapor disappears, which leads to a significant increase in the absorption of laser radiation~\cite{2017CoPhC.214...59B}. The space around the target is filled with vaporized matter, which continues to heat up and ionizes, and a hot emitting plasma is formed. In the case ``We = 20'', in the region near the critical surface ($n_e$ = 10$^{21}$ cm$^{-3}$) with the side of the laser irradiated, the plasma reaches a temperature of 4 eV (see Fig.~\ref{temperature_20ns}) at the time of the maximum laser pulse ($t_0$ = 20 ns).

Such a hot plasma scatters at high velocity, and its total momentum is directed toward the laser. In this case, the target receives a recoil momentum in accordance with the law of conservation of momentum and begins to move at an acquired velocity that increases during the laser impact. After the laser is turned off, the target velocity remains constant. According to the RALEF-2D code, it is 7.54 m/s, which is in good agreement with the data obtained in experiment, namely 7.99 m/s for the case ``We = 20''~\cite{Kurilovich2018}. The fraction of the substance vaporized by the laser is only 0.35 $\pm$ 0.05 \% of the initial mass of the droplet, however, this is sufficient for the generated hot plasma to give the target a significant impulse.

\begin{figure}[!htb]
    \centering
\begingroup
  \makeatletter
  \providecommand\color[2][]{%
    \GenericError{(gnuplot) \space\space\space\@spaces}{%
      Package color not loaded in conjunction with
      terminal option `colourtext'%
    }{See the gnuplot documentation for explanation.%
    }{Either use 'blacktext' in gnuplot or load the package
      color.sty in LaTeX.}%
    \renewcommand\color[2][]{}%
  }%
  \providecommand\includegraphics[2][]{%
    \GenericError{(gnuplot) \space\space\space\@spaces}{%
      Package graphicx or graphics not loaded%
    }{See the gnuplot documentation for explanation.%
    }{The gnuplot epslatex terminal needs graphicx.sty or graphics.sty.}%
    \renewcommand\includegraphics[2][]{}%
  }%
  \providecommand\rotatebox[2]{#2}%
  \@ifundefined{ifGPcolor}{%
    \newif\ifGPcolor
    \GPcolortrue
  }{}%
  \@ifundefined{ifGPblacktext}{%
    \newif\ifGPblacktext
    \GPblacktexttrue
  }{}%
  \let\gplgaddtomacro\g@addto@macro
  \gdef\gplbacktext{}%
  \gdef\gplfronttext{}%
  \makeatother
  \ifGPblacktext
    \def\colorrgb#1{}%
    \def\colorgray#1{}%
  \else
    \ifGPcolor
      \def\colorrgb#1{\color[rgb]{#1}}%
      \def\colorgray#1{\color[gray]{#1}}%
      \expandafter\def\csname LTw\endcsname{\color{white}}%
      \expandafter\def\csname LTb\endcsname{\color{black}}%
      \expandafter\def\csname LTa\endcsname{\color{black}}%
      \expandafter\def\csname LT0\endcsname{\color[rgb]{1,0,0}}%
      \expandafter\def\csname LT1\endcsname{\color[rgb]{0,1,0}}%
      \expandafter\def\csname LT2\endcsname{\color[rgb]{0,0,1}}%
      \expandafter\def\csname LT3\endcsname{\color[rgb]{1,0,1}}%
      \expandafter\def\csname LT4\endcsname{\color[rgb]{0,1,1}}%
      \expandafter\def\csname LT5\endcsname{\color[rgb]{1,1,0}}%
      \expandafter\def\csname LT6\endcsname{\color[rgb]{0,0,0}}%
      \expandafter\def\csname LT7\endcsname{\color[rgb]{1,0.3,0}}%
      \expandafter\def\csname LT8\endcsname{\color[rgb]{0.5,0.5,0.5}}%
    \else
      \def\colorrgb#1{\color{black}}%
      \def\colorgray#1{\color[gray]{#1}}%
      \expandafter\def\csname LTw\endcsname{\color{white}}%
      \expandafter\def\csname LTb\endcsname{\color{black}}%
      \expandafter\def\csname LTa\endcsname{\color{black}}%
      \expandafter\def\csname LT0\endcsname{\color{black}}%
      \expandafter\def\csname LT1\endcsname{\color{black}}%
      \expandafter\def\csname LT2\endcsname{\color{black}}%
      \expandafter\def\csname LT3\endcsname{\color{black}}%
      \expandafter\def\csname LT4\endcsname{\color{black}}%
      \expandafter\def\csname LT5\endcsname{\color{black}}%
      \expandafter\def\csname LT6\endcsname{\color{black}}%
      \expandafter\def\csname LT7\endcsname{\color{black}}%
      \expandafter\def\csname LT8\endcsname{\color{black}}%
    \fi
  \fi
    \setlength{\unitlength}{0.0500bp}%
    \ifx\gptboxheight\undefined%
      \newlength{\gptboxheight}%
      \newlength{\gptboxwidth}%
      \newsavebox{\gptboxtext}%
    \fi%
    \setlength{\fboxrule}{0.5pt}%
    \setlength{\fboxsep}{1pt}%
\begin{picture}(5102.00,3400.00)%
    \gplgaddtomacro\gplbacktext{%
      \csname LTb\endcsname
      \put(814,704){\makebox(0,0)[r]{\strut{}$0$}}%
      \csname LTb\endcsname
      \put(814,1199){\makebox(0,0)[r]{\strut{}$0.5$}}%
      \csname LTb\endcsname
      \put(814,1694){\makebox(0,0)[r]{\strut{}$1$}}%
      \csname LTb\endcsname
      \put(814,2189){\makebox(0,0)[r]{\strut{}$1.5$}}%
      \csname LTb\endcsname
      \put(814,2684){\makebox(0,0)[r]{\strut{}$2$}}%
      \csname LTb\endcsname
      \put(814,3179){\makebox(0,0)[r]{\strut{}$2.5$}}%
      \csname LTb\endcsname
      \put(946,484){\makebox(0,0){\strut{}$0$}}%
      \csname LTb\endcsname
      \put(1886,484){\makebox(0,0){\strut{}$0.5$}}%
      \csname LTb\endcsname
      \put(2826,484){\makebox(0,0){\strut{}$1$}}%
      \csname LTb\endcsname
      \put(3765,484){\makebox(0,0){\strut{}$1.5$}}%
      \csname LTb\endcsname
      \put(4705,484){\makebox(0,0){\strut{}$2$}}%
    }%
    \gplgaddtomacro\gplfronttext{%
      \csname LTb\endcsname
      \put(209,1941){\rotatebox{-270}{\makebox(0,0){\strut{}Pressure, kbar}}}%
      \put(2825,154){\makebox(0,0){\strut{}Angle, rad}}%
    }%
    \gplbacktext
    \put(0,0){\includegraphics{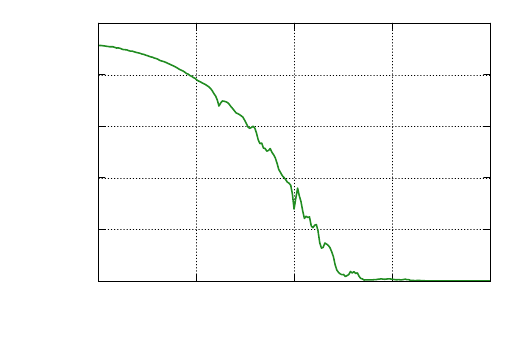}}%
    \gplfronttext
  \end{picture}%
\endgroup
    \caption{Pressure on the irradiated surface of the target.}
    \label{pressure_20ns}
\end{figure}

\begin{figure}[!htb]
    \centering
    \includegraphics[width=0.8\linewidth]{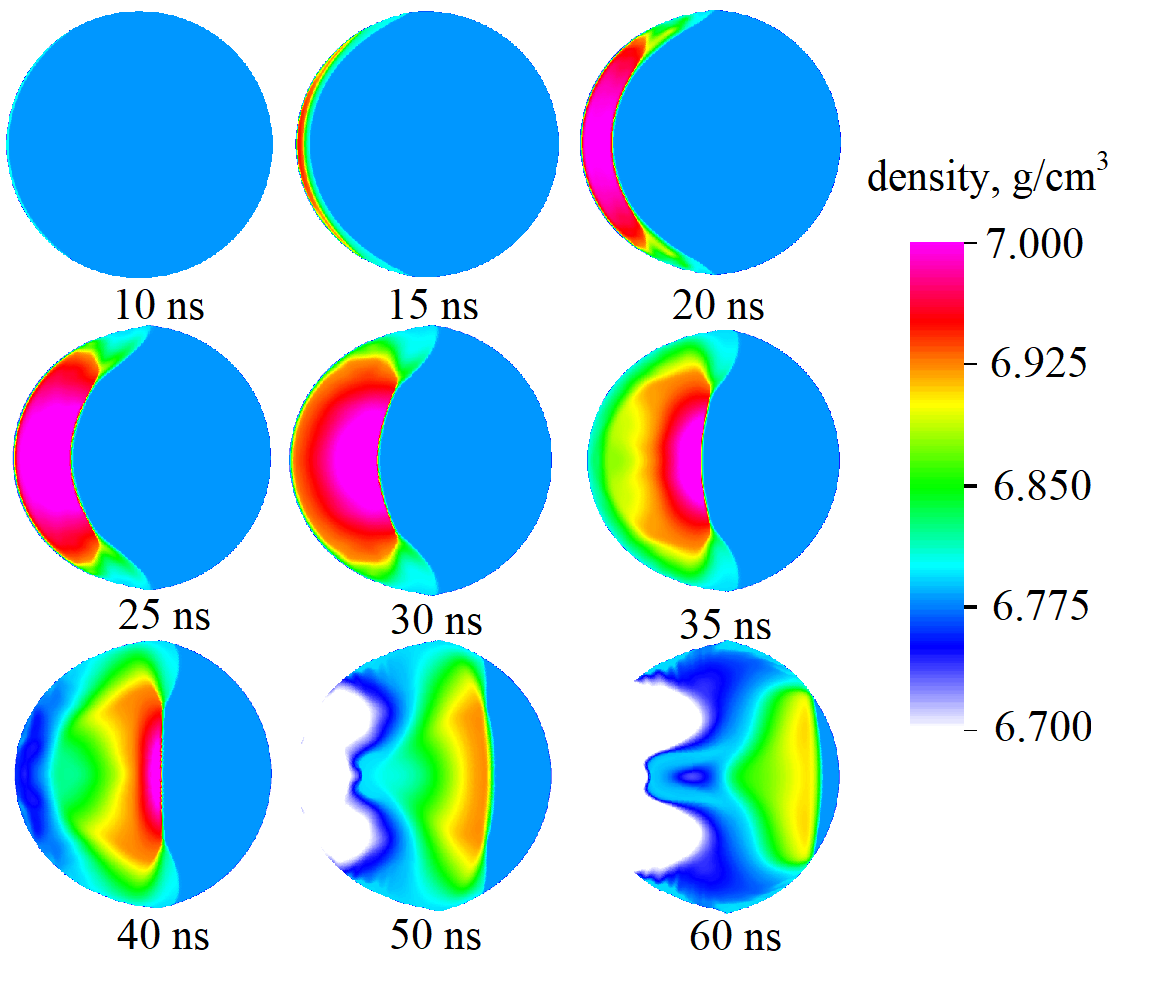}
    \caption{Shock wave propagation through a tin drop after laser pulse impact.}
    \label{Shock}
\end{figure}

A part of the received energy dissipates inside the target and also transfers into the kinetic energy of deformation. The pressure at the hemispherical surface of the droplet closest to the laser reaches 2~kbar and has an angular distribution depending on the spatial profile of the laser beam and the incidence angle $\theta$ counted from the central axis, see Fig.~\ref{pressure_20ns}. Such a spatial distribution of pressure leads to the formation of a hemispherical weak shock wave, which begins to propagate to the center of the target with an almost sound velocity~\cite{Zeldovich2002}. Converging to the center of the target, the amplitude of the shock wave first attenuates slowly, and then slightly increases due to hydrodynamic focusing. In the case of high-intensity picosecond laser pulses~\cite{Basko2017}, the focusing effect led to the formation of an extensive resonator in the center of the target. Here, this effect is small, since the amplitude of the shock wave is three orders of magnitude smaller, an extensive cavity does not arise, but some density inhomogeneity appears. It is also important to note that rarefaction waves arise after reflection from the free boundary. Moreover, they come from different parts of the spherical surface with different amplitudes. Since the laser exposure time in the cases under consideration is comparable with the wave propagation time through the target ($\sim$60~ns), there are multiple imposition of rarefaction waves reflected from the spherical surface on top of each other and new compression waves caused in the already rarefied substance by the laser that continues to impact. The interference of these waves inside the target leads to a complex picture of local density inhomogeneities. After the passage of the shock wave, the target substance turns into a homogeneous mixture of liquid and vapor~\cite{Basko_developmentof, Basko2012}. In the P-V phase diagram, this process is expressed by the intersection of the unloading isentropes and the binodal, which leads to a discontinuity in the speed of sound. A discontinuity in the speed of sound leads to compaction/rarefaction of the substance~\cite{Zeldovich2002, Basko2017, Inogamov2016}. Some numerical perturbations due to a discrete tabular representation of the equation of state are imposed on this process, which leads to an additional numerical discontinuity in the speed of sound at some points of the computational domain behind the shock wavefront.

In general, it can be said that a combination of several factors affects the inhomogeneous distribution of the density inside the target after the passage of a weak shock wave, namely:
\begin{itemize}
    \item spherical geometry (converging to the center of a spherical target, the shock wave changes its amplitude, attenuating and somewhat amplifying in the center due to hydrodynamic focusing);
    \item the duration of the laser pulse comparable with the characteristic hydrodynamic times of passage of the shock wave through the target;
    \item interference of rarefaction waves coming from different parts of the free boundary;
    \item discontinuity in the speed of sound after the passage of the shock wave;
    \item some numerical perturbation associated with the discreteness of the tabular equation of state.
\end{itemize}

As a result, after the end of laser impact and the passage of the shock wave, the target is an almost spherical object, consisting of a homogeneous mixture of liquid and vapor, with a density close to the initial one. In this case, there are numerous local areas of compaction/rarefaction of the substance throughout the target volume, which in the subsequent modeling of the second stage with allowance for surface tension forces leads to the formation of gas bubbles.

The spatial distribution of pressure and velocity field inside the target after exposure to a laser pulse is such that the drop gradually flattened. At time 100~ns, the dimensions of the target acquire an aspect ratio of the diameter (maximum dimension along the r axis) to thickness (maximum dimension along the Z-axis) of the order of 52$\div$50.4. Further, over time, due to inertia forces, this aspect ratio continues to change in the direction of increasing diameter in relation to the thickness, the target deforms into a disk. The processes of deformation of water droplets after exposure by laser pulse are described in sufficient detail~\cite{gelderblom_lhuissier_klein_bouwhuis_lohse_villermaux_snoeijer_2016}. In the case of LMD, the deformation process can be described similarly.

The presented results of numerical simulation of the first stage were obtained using the RALEF-2D code. It was developed on the basis of a self-consistent model, which includes two-dimensional hydro- and gasdynamics, radiation transfer, thermal conductivity, and the interaction of laser radiation with matter, taking into account refraction. In RALEF-2D code, to solve the equations of gas dynamics, a Lagrangian-Eulerian numerical scheme of the second-order of the Godunov type on adaptive grids is implemented~\cite{Addessio:1992fh}.

\begin{figure}[!htb]
    \centering
    \includegraphics[width=0.9\linewidth]{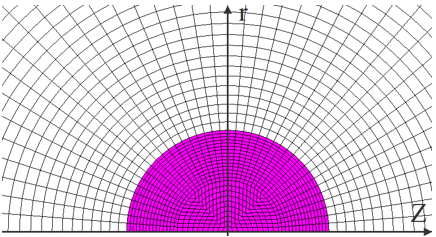}
    \caption{Schematic representation of the computational mesh used in RALEF-2D for numerical simulation of the problem under consideration.}
    \label{Grid_ralef}
\end{figure}

In calculations, we used the nonuniform mesh with cell densification in the target region, the minimum size of which is about 80~nm at the boundary between the target and the surrounding gas. The schematic representation of the computational mesh used in simulations is shown in Fig.~\ref{Grid_ralef}.

The radiation transfer equation is solved by the Sn-method. The absorption coefficients included in the radiation transfer equation were obtained using the THERMOS~\cite{Novikov2016} in the approximation of collisional-radiative equilibrium. In calculating the absorption coefficients, the Planck radiation field in the EUV range is used, where the effects of photon reabsorption are significant. Outside this range, the radiation field was assumed to be zero. A distinctive feature of these RALEF-2D calculations is the use of the two-phase equation of state for tin, which allows to describe the liquid-vapor phase transition. The equation of state was calculated according to the FEOS model in the equilibrium approximation with the Maxwellian construction~\cite{Faik2012}.

Unfortunately, to further calculations of the target deformation after turning off the laser, the described model without surface tension forces is not applicable. If we continue the simulation using RALEF-2D code, the target will expand infinitely with time, and the density will be diffusion spread over space without a sharp liquid–vapor interface (see Fig.~\ref{Ralef_OF_We20}). But this behavior is unphysical. Firstly, the liquid target after cooling has a clear boundary, and secondly, its radial expansion will be inhibited due to surface tension forces, and at a certain point it will begin to contract back into a spherical drop. Therefore, to continue modeling the second stage, it becomes necessary to use another model.

\subsection{Stage 2: deformation of freely moving target}
\label{subSection_Stage2}

A few tens of nanoseconds after the end of the laser pulse, the second stage of the target deformation begins. The target cools quickly enough, and the plasma that has flown to the periphery no longer has a significant effect on it. For further modeling, it is no longer necessary to take into account radiation processes and thermal conductivity. The target moves with an acquired constant speed by inertia and expands, tending to the shape of the disk, while the radial expansion velocity strongly depends on the surface tension forces.

To calculate the deformation of the target, we used the volume of fluid (VOF) method~\cite{Hirt1981} as part of the Finite Volume Method~\cite{Moukalled}, implemented in the \texttt{interFoam}/ \texttt{interDyMFoam} solver for two-phase incompressible, immiscible, and isothermal fluids. The contribution of surface tension forces is taken into account according to the continuum surface force (CSF) model~\cite{Brackbill1992}. To solve this problem, an incompressible fluid model is applicable, since the speed of sound in a substance (the speed of sound in liquid tin is about 2.5~km/s) is much higher than the speed of motion of the medium itself (10~--~100~m/s). The possibility of using an incompressible fluid model greatly simplifies the solution of the problem.

The method of the end-to-end simulation of the target deformation proposed by the authors is based on the task of transferring the data calculated using RALEF-2D as initial conditions for the subsequent simulation of the problem by the OpenFOAM code. It is very important to choose the correct moment in time at which the transition from one model to another takes place. After turning off the laser, for some time a cloud of plasma and hot gas exists near the target, and the surface of the target itself has a temperature near or slightly above the boiling point. While the target is hot enough, active evaporation from the surface occurs. Obviously, radiation transfer and thermal conductivity cannot be neglected at this time, and the temperature regime and the evaporation process require the use of an adequate equation of state. Therefore, the simulation using RALEF-2D must be continued for some time after the end of the laser pulse, while the drop cools. In this case, the process of the deformation is already underway. Indeed, if the surface temperature of a liquid body is lower than the boiling point, there is a sharp boundary between the liquid and the vapor, determined by the surface tension forces. However, the RALEF-2D model provides a relatively smooth density gradient instead of a sharp boundary even at temperatures below the boiling point, which in fact leads to fictitious evaporation and the formation of an unnatural target shape. The optimal time interval between the moment of switching off the laser pulse and the transition to modeling the deformation of the target turned out to be a value lying within 40~--~80~ns. A time moment of 100~ns (60~ns after the end of the laser pulse) was selected. By this moment, the cloud of hot plasma has scattered, and around the target in the region with a characteristic size of the order of 10 of its radii, there is only a low-density ($\sim$~10$^{-5}$~g/cm$^3$) gas with a temperature below 0.5~eV. The surface of the liquid target has cooled to 1700 K, which is lower than the boiling point. In turn, the shape of the target is still not very different from the initial drop and the effect of the “diffusion smeared boundary” is not large.

\begin{figure}[!htb]
    \centering
    \includegraphics[width=0.5\linewidth]{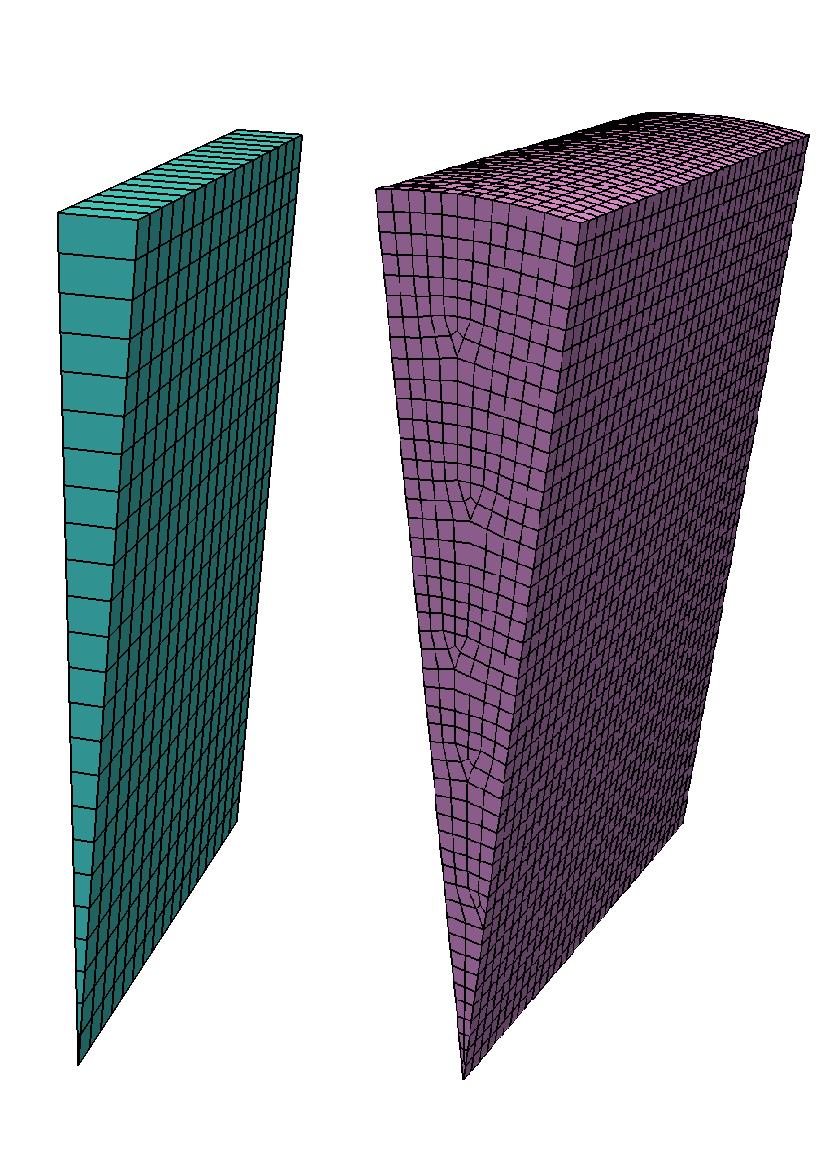}
    \caption{Schematic representation of computational grids used in OpenFOAM for modeling the problem under consideration. Single-layer mesh for 2D-simulation (left) and multi-layer for 3D-simulation (right).}
    \label{Grid_of}
\end{figure}

The OpenFOAM simulation was carried out in the 2D and 3D versions. In this case, under 2D is understood as the fact that the simulation is carried out on an axisymmetric single-layer (one cell in the layer)  of the computational mesh uniform in radius. The mesh cells are three-dimensional, and the three-dimensional equations of hydrodynamics with periodic boundary conditions at the corresponding boundaries are solved in them. In 3D simulation, a similar multilayer mesh is used. Geometrically, both meshes represent a sector of the cylinder with a sweep angle of 5$^{\circ}$ in the case of 2D and 15$^{\circ}$ in the case of 3D (see Fig.~\ref{Grid_of}). The 3D simulation on the 30$^{\circ}$ also has been performed, however, no changes in the target shape were found. The choice of computational mesh is due to the need for fairly accurate simulation in a reasonable time.


Since it is required to transfer the distribution of quantities from the 2D axisymmetric mesh (r, z) (see Fig.~\ref{Grid_ralef}) to the 3D in the Cartesian coordinates (x, y, z) (see Fig.~\ref{Grid_of}), we first perform the transformation of the axisymmetric mesh in volume (by rotation around the axis of symmetry).

\begin{figure}[!htb]
    \centering
    \includegraphics[width=0.9\linewidth]{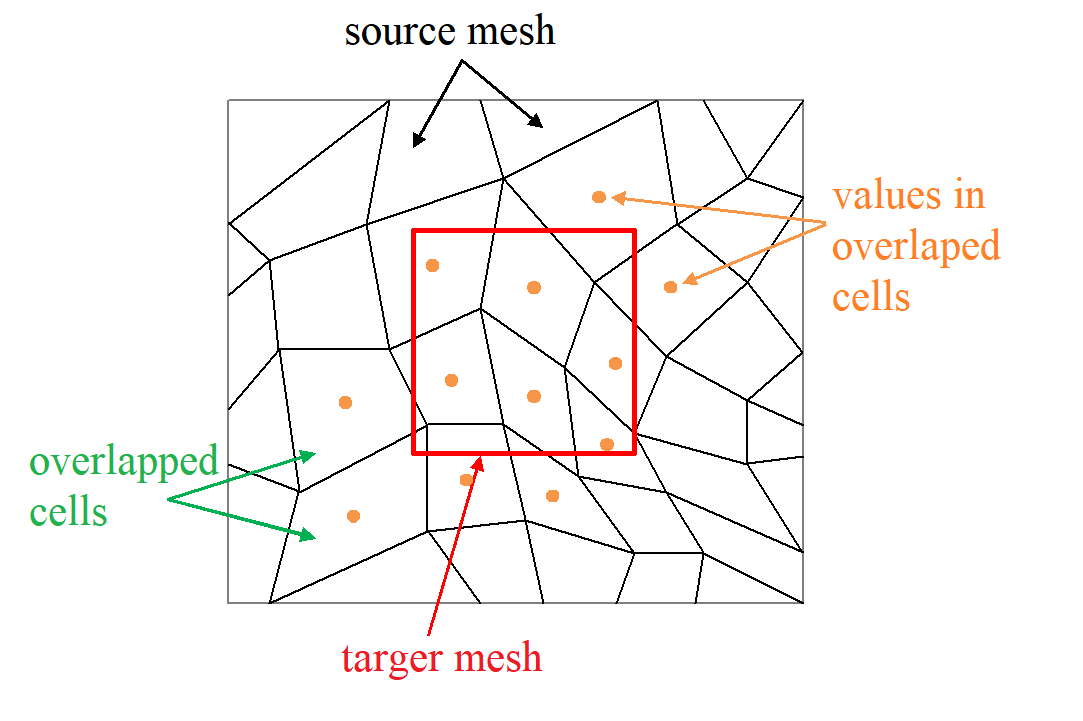}
    \caption{Schematic representation of intersections of the cell volume (target mesh) with the original mesh.}
    \label{cvw}
\end{figure}

When transferring density and velocity fields from the source mesh to the target mesh, the cell volume weighted (CVW) method of interpolation is used. The value of a cell weighted by volume is calculated from the values of the cells of the source mesh overlapped by the cell of the target mesh, as well as the volume fractions of the overlaps between the cells. Figure~\ref{cvw} shows the intersection areas of the overlapped cells of the source mesh with the cell of the target mesh.

The value of $\varphi$ in the cell of target mesh is determined by the following formula
\begin{equation}
\varphi_{t}=\frac{1}{v_{t}} \sum_{i=1}^{n} v_{int; i} \cdot \varphi_{s; i},
\end{equation}
where $v_t$ is the volume of the cell of the target mesh, n is the number of overlapped cells of the source mesh with the cell of the target mesh with the overlap volume $v_{int}$ other than zero, $\varphi_s$ is the value of $\varphi$ in the center of the cell of the source mesh.

Calculation of cut off volumes is quite a time-consuming procedure, but very high-quality interpolation is obtained. The most important advantage of volume-weighted interpolation is that the interpolation of values between overlapping meshes is conservative.

The computational mesh used in OpenFOAM is quite detailed, which is associated with the need for good resolution in the region of the liquid and the surrounding gas interface. The typical cell size of the mesh in OpenFOAM simulation is 100~nm for 2D and 250~nm for 3D.

\begin{figure}[!htb]
    \centering
    \includegraphics[width=0.9\linewidth]{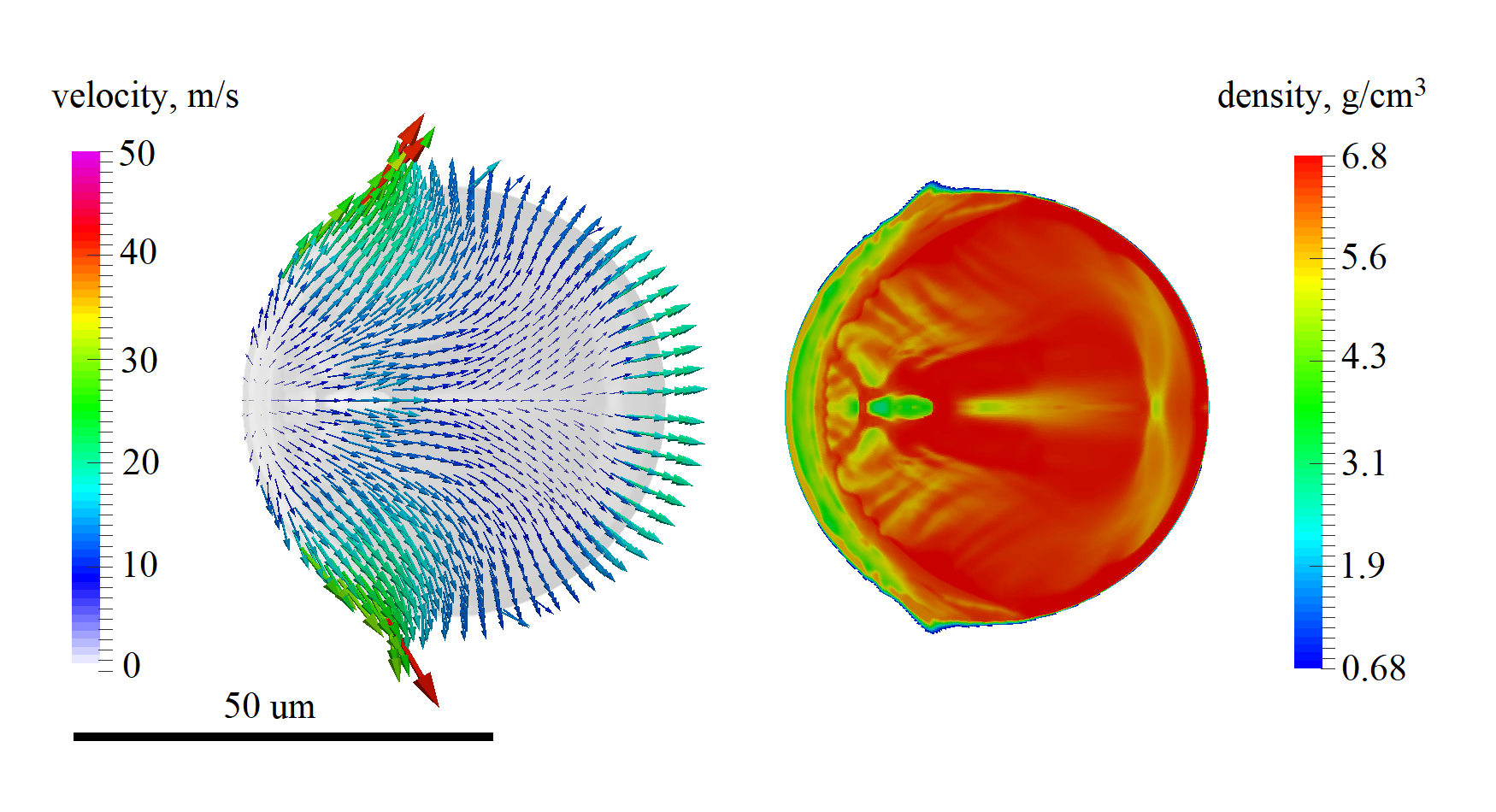}
    \caption{Velocity (left) and density (right) distributions inside the target at a time of 100 ns, obtained using RALEF-2D and transferred to OpenFOAM for further simulation of the deformation process.}
    \label{Fields_We20}
\end{figure}

Since the rarefied gas surrounding the target has practically no effect on its deformation, it is excluded from the simulation. In OpenFOAM only density is taken into account in the range from 10 to 100 \% of the maximum density, i.e. from 0.68 to 6.8~g/cm$^3$. Taking into account the fact that evaporation was only 0.35~$\pm$~0.05 \% of the initial mass of the droplet, the calculation error as a result of this action is negligible. The transferred velocity and density fields from the RALEF-2D simulation at a time instant of 100 ns are shown in Fig.~\ref{Fields_We20}.

When switching to simulation using OpenFOAM, the computational domain was reduced to sizes of 60$\times$80~$\mu$m (60~$\mu$m along the Z-axis and 80~$\mu$m along the radius). The total number of mesh cells reached 480000 in the case of 2D and 3 million for 3D, which is significantly more than in the simulation of RALEF-2D (just over 150000), where a non-uniform mesh is used.

\section{Simulation results and comparison with experiment}
\label{section_Results}

\begin{figure}[!htb]
    \centering
\begingroup
  \makeatletter
  \providecommand\color[2][]{%
    \GenericError{(gnuplot) \space\space\space\@spaces}{%
      Package color not loaded in conjunction with
      terminal option `colourtext'%
    }{See the gnuplot documentation for explanation.%
    }{Either use 'blacktext' in gnuplot or load the package
      color.sty in LaTeX.}%
    \renewcommand\color[2][]{}%
  }%
  \providecommand\includegraphics[2][]{%
    \GenericError{(gnuplot) \space\space\space\@spaces}{%
      Package graphicx or graphics not loaded%
    }{See the gnuplot documentation for explanation.%
    }{The gnuplot epslatex terminal needs graphicx.sty or graphics.sty.}%
    \renewcommand\includegraphics[2][]{}%
  }%
  \providecommand\rotatebox[2]{#2}%
  \@ifundefined{ifGPcolor}{%
    \newif\ifGPcolor
    \GPcolortrue
  }{}%
  \@ifundefined{ifGPblacktext}{%
    \newif\ifGPblacktext
    \GPblacktexttrue
  }{}%
  \let\gplgaddtomacro\g@addto@macro
  \gdef\gplbacktext{}%
  \gdef\gplfronttext{}%
  \makeatother
  \ifGPblacktext
    \def\colorrgb#1{}%
    \def\colorgray#1{}%
  \else
    \ifGPcolor
      \def\colorrgb#1{\color[rgb]{#1}}%
      \def\colorgray#1{\color[gray]{#1}}%
      \expandafter\def\csname LTw\endcsname{\color{white}}%
      \expandafter\def\csname LTb\endcsname{\color{black}}%
      \expandafter\def\csname LTa\endcsname{\color{black}}%
      \expandafter\def\csname LT0\endcsname{\color[rgb]{1,0,0}}%
      \expandafter\def\csname LT1\endcsname{\color[rgb]{0,1,0}}%
      \expandafter\def\csname LT2\endcsname{\color[rgb]{0,0,1}}%
      \expandafter\def\csname LT3\endcsname{\color[rgb]{1,0,1}}%
      \expandafter\def\csname LT4\endcsname{\color[rgb]{0,1,1}}%
      \expandafter\def\csname LT5\endcsname{\color[rgb]{1,1,0}}%
      \expandafter\def\csname LT6\endcsname{\color[rgb]{0,0,0}}%
      \expandafter\def\csname LT7\endcsname{\color[rgb]{1,0.3,0}}%
      \expandafter\def\csname LT8\endcsname{\color[rgb]{0.5,0.5,0.5}}%
    \else
      \def\colorrgb#1{\color{black}}%
      \def\colorgray#1{\color[gray]{#1}}%
      \expandafter\def\csname LTw\endcsname{\color{white}}%
      \expandafter\def\csname LTb\endcsname{\color{black}}%
      \expandafter\def\csname LTa\endcsname{\color{black}}%
      \expandafter\def\csname LT0\endcsname{\color{black}}%
      \expandafter\def\csname LT1\endcsname{\color{black}}%
      \expandafter\def\csname LT2\endcsname{\color{black}}%
      \expandafter\def\csname LT3\endcsname{\color{black}}%
      \expandafter\def\csname LT4\endcsname{\color{black}}%
      \expandafter\def\csname LT5\endcsname{\color{black}}%
      \expandafter\def\csname LT6\endcsname{\color{black}}%
      \expandafter\def\csname LT7\endcsname{\color{black}}%
      \expandafter\def\csname LT8\endcsname{\color{black}}%
    \fi
  \fi
    \setlength{\unitlength}{0.0500bp}%
    \ifx\gptboxheight\undefined%
      \newlength{\gptboxheight}%
      \newlength{\gptboxwidth}%
      \newsavebox{\gptboxtext}%
    \fi%
    \setlength{\fboxrule}{0.5pt}%
    \setlength{\fboxsep}{1pt}%
\begin{picture}(5102.00,3400.00)%
    \gplgaddtomacro\gplbacktext{%
      \csname LTb\endcsname
      \put(814,1275){\makebox(0,0)[r]{\strut{}$0$}}%
      \csname LTb\endcsname
      \put(814,1656){\makebox(0,0)[r]{\strut{}$20$}}%
      \csname LTb\endcsname
      \put(814,2037){\makebox(0,0)[r]{\strut{}$40$}}%
      \csname LTb\endcsname
      \put(814,2417){\makebox(0,0)[r]{\strut{}$60$}}%
      \csname LTb\endcsname
      \put(814,2798){\makebox(0,0)[r]{\strut{}$80$}}%
      \csname LTb\endcsname
      \put(814,3179){\makebox(0,0)[r]{\strut{}$100$}}%
      \csname LTb\endcsname
      \put(946,484){\makebox(0,0){\strut{}$0$}}%
      \csname LTb\endcsname
      \put(1483,484){\makebox(0,0){\strut{}$2$}}%
      \csname LTb\endcsname
      \put(2020,484){\makebox(0,0){\strut{}$4$}}%
      \csname LTb\endcsname
      \put(2557,484){\makebox(0,0){\strut{}$6$}}%
      \csname LTb\endcsname
      \put(3094,484){\makebox(0,0){\strut{}$8$}}%
      \csname LTb\endcsname
      \put(3631,484){\makebox(0,0){\strut{}$10$}}%
      \csname LTb\endcsname
      \put(4168,484){\makebox(0,0){\strut{}$12$}}%
      \csname LTb\endcsname
      \put(4705,484){\makebox(0,0){\strut{}$14$}}%
    }%
    \gplgaddtomacro\gplfronttext{%
      \csname LTb\endcsname
      \put(209,1941){\rotatebox{-270}{\makebox(0,0){\strut{}radius, $\mu$m}}}%
      \put(2825,154){\makebox(0,0){\strut{}time, $\mu$s}}%
      \csname LTb\endcsname
      \put(3718,1757){\makebox(0,0)[r]{\strut{}We = 20}}%
      \csname LTb\endcsname
      \put(3718,1537){\makebox(0,0)[r]{\strut{}RALEF 0.5 g/cm$^3$}}%
      \csname LTb\endcsname
      \put(3718,1317){\makebox(0,0)[r]{\strut{}2D}}%
      \csname LTb\endcsname
      \put(3718,1097){\makebox(0,0)[r]{\strut{}3D}}%
      \csname LTb\endcsname
      \put(3718,877){\makebox(0,0)[r]{\strut{}2D (uniform density)}}%
    }%
    \gplbacktext
    \put(0,0){\includegraphics{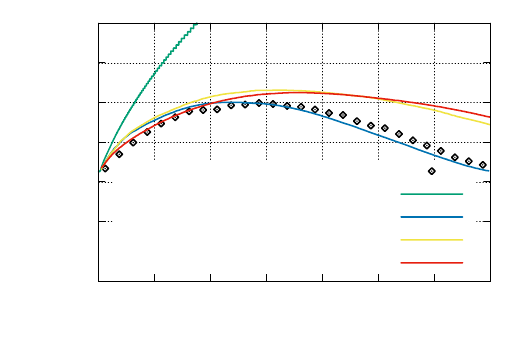}}%
    \gplfronttext
  \end{picture}%
\endgroup
    \caption{Target radius evolution over time. ARCNL We20 experiment (black dots) and calculations (color curves).}
    \label{Radii_We20}
\end{figure}

\begin{figure}[!htb]
    \centering
    \includegraphics[width=0.9\linewidth]{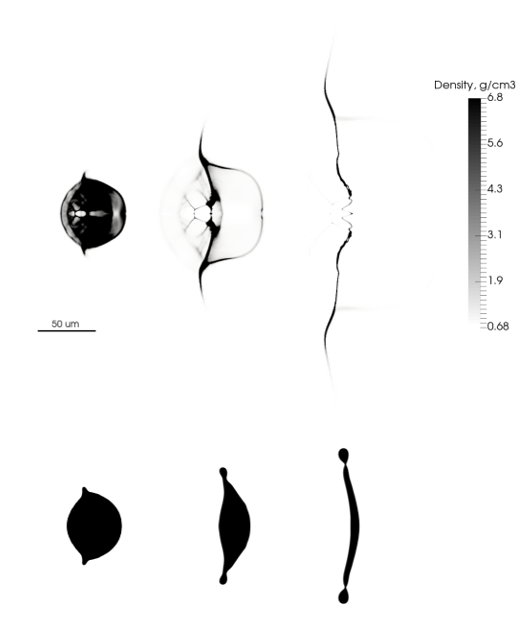}
    \caption{Target deformation after exposure to a laser pulse. Simulation using RALEF-2D without taking into account surface tension (top). The result of joint modeling using RALEF-2D and OpenFOAM codes, 2D simulation with uniform density (bottom).}
    \label{Ralef_OF_We20}
\end{figure}

\begin{figure}[!htb]
    \centering
    \includegraphics[width=0.9\linewidth]{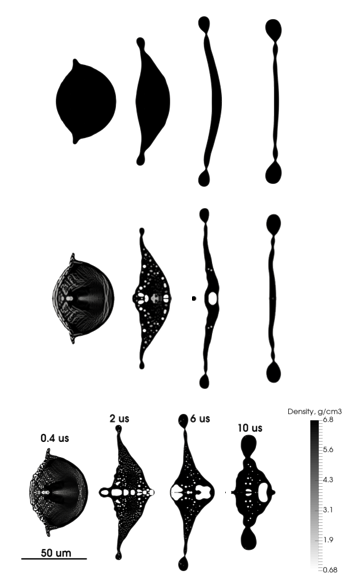}
    \caption{Target deformation after exposure by laser pulse. The result of joint modeling using RALEF-2D and OpenFOAM codes. 2D simulation (bottom), 2D simulation with uniform density (top), 3D simulation (middle).}
    \label{OF_We20}
\end{figure}

To study the LMD deformation after exposure by a laser pulse, four modeling options were tested. In the first case, the RALEF-2D simulation was used not only for the first stage of the process but also for the second, that is, without taking into account surface tension forces. The other three modeling options consisted of a joint sequential simulation using RALEF-2D for the first stage of the process and OpenFOAM for the second stage, where the 3D analysis, or 2D simulation, or 2D simulation with uniform density were performed using OpenFOAM. After turning off the laser, the target has a large number of local density inhomogeneities, which are converted into gas bubbles, after transferring data to OpenFOAM and starting the simulation with surface tension forces. It turned out that the presence of these bubbles significantly affects the process of further deformation of the target. In order to better understand this phenomenon, which has both a numerical and possibly real physical nature, a variant of 2D simulation with a uniform density was added to the analysis, in which, when transferring data from RALEF-2D to OpenFOAM, the density was artificially smoothed inside the contour of the target (in compliance with the laws of mass and momentum conservation). By the way, the simulation of 2D with a uniform density is the least expensive in terms of the use of computational resources, since the effective surface of the liquid in the absence of bubbles is much smaller. The results obtained for the dependencies of the target radius on time are shown in Fig.~\ref{Radii_We20} together with experimental data for the case ``We = 20''~\cite{PhysRevApplied.6.014018}.
Figure~\ref{Ralef_OF_We20} shows the target deformation, where the upper part shows the result obtained using only RALEF-2D code, and the lower part shows the combined simulation of RALEF-2D and OpenFOAM (2D variant with uniform density). A comparison of the three options for the joint simulation of the target deformation is presented in Fig.~\ref{OF_We20}. In Fig.~\ref{Ralef_OF_We20} for the first variant, where both stages were calculated using RALEF-2D, a curve is presented that corresponds to a density contour of 0.5~g/cm$^3$. Since in this simulation option there is no clear boundary between the liquid and the vapor, it is necessary to choose a certain threshold density value for it. It is assumed that a density of 0.5~g/cm$^3$ can still be seen in experimental images of the target, with which the radius is determined.

Figure~\ref{Radii_We20} shows that the simulation option without taking into account the surface tension forces is in poor agreement with the experiment, the expansion of the target proceeds faster, and the curve of the radius versus time do not have an extremum corresponding to the moment when the liquid disk is starting to return into the drop. The other three colored curves related to the joint simulation with taking into account surface tension forces lie much closer to the experimental curve and have an extremum. Thus, the first conclusion that can be drawn is the correctness of the chosen approach -- to use the joint simulation of two codes with different models, each of which is most applicable for the corresponding stage of the deformation process. Let's discuss in more detail three colored curves in Fig.~\ref{Radii_We20}, corresponding to the joint simulation. Two facts should be highlighted here, the first -- the 3D simulation almost coincided with the 2D simulation with a uniform density, the second -- the 2D simulation with the density field transmitted without artificial manipulations, better than the others, matches with the experiment. {\color{red}In the 2D simulation, bubbles are hollow tori in three-dimensional geometry, which, due to surface tension forces, tend to contract to the axis, dragging the surrounding mass. This effect leads to a faster compression of the target back into the drop, which is observed in Fig.~\ref{Radii_We20} (the blue curve deviates from red and yellow).}

\begin{figure}[!htb]
    \centering
    \includegraphics[width=0.5\linewidth]{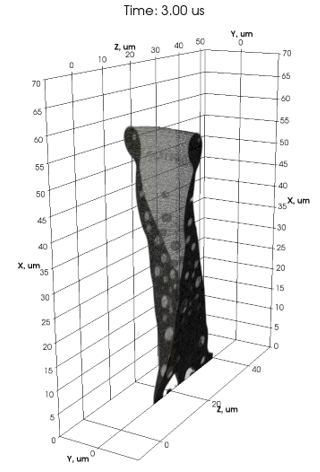}
    \caption{Three-dimensional image of the target deformation after exposure by laser pulse.}
    \label{OF_3D_We20}
\end{figure}

In the 3D case, the tori decay, turning into spherical gas bubbles, and their position in space changes only due to the hydrodynamic flow of the liquid (see Fig.~\ref{OF_3D_We20}).

So, the second conclusion that can be drawn from the results of the study is the following: that of the three options for joint simulation, the best agreement with the experiment gives the 2D option, taking into account density inhomogeneities, is most likely an accident. Authors are more likely to trust the results of 3D and 2D calculations with uniform density, since it is intuitively hard to believe that toroidal gas formations can stably exist inside a liquid tin disk.

The work done has fully proved the applicability of the proposed methodology for end-to-end modeling of LMD deformation after exposure by a laser pulse. The simulations showed that for better agreement with the experiment, it is necessary to clarify a number of physical and numerical parameters embedded in the models used.

\begin{figure}[!htb]
    \centering
\begingroup
  \makeatletter
  \providecommand\color[2][]{%
    \GenericError{(gnuplot) \space\space\space\@spaces}{%
      Package color not loaded in conjunction with
      terminal option `colourtext'%
    }{See the gnuplot documentation for explanation.%
    }{Either use 'blacktext' in gnuplot or load the package
      color.sty in LaTeX.}%
    \renewcommand\color[2][]{}%
  }%
  \providecommand\includegraphics[2][]{%
    \GenericError{(gnuplot) \space\space\space\@spaces}{%
      Package graphicx or graphics not loaded%
    }{See the gnuplot documentation for explanation.%
    }{The gnuplot epslatex terminal needs graphicx.sty or graphics.sty.}%
    \renewcommand\includegraphics[2][]{}%
  }%
  \providecommand\rotatebox[2]{#2}%
  \@ifundefined{ifGPcolor}{%
    \newif\ifGPcolor
    \GPcolortrue
  }{}%
  \@ifundefined{ifGPblacktext}{%
    \newif\ifGPblacktext
    \GPblacktexttrue
  }{}%
  \let\gplgaddtomacro\g@addto@macro
  \gdef\gplbacktext{}%
  \gdef\gplfronttext{}%
  \makeatother
  \ifGPblacktext
    \def\colorrgb#1{}%
    \def\colorgray#1{}%
  \else
    \ifGPcolor
      \def\colorrgb#1{\color[rgb]{#1}}%
      \def\colorgray#1{\color[gray]{#1}}%
      \expandafter\def\csname LTw\endcsname{\color{white}}%
      \expandafter\def\csname LTb\endcsname{\color{black}}%
      \expandafter\def\csname LTa\endcsname{\color{black}}%
      \expandafter\def\csname LT0\endcsname{\color[rgb]{1,0,0}}%
      \expandafter\def\csname LT1\endcsname{\color[rgb]{0,1,0}}%
      \expandafter\def\csname LT2\endcsname{\color[rgb]{0,0,1}}%
      \expandafter\def\csname LT3\endcsname{\color[rgb]{1,0,1}}%
      \expandafter\def\csname LT4\endcsname{\color[rgb]{0,1,1}}%
      \expandafter\def\csname LT5\endcsname{\color[rgb]{1,1,0}}%
      \expandafter\def\csname LT6\endcsname{\color[rgb]{0,0,0}}%
      \expandafter\def\csname LT7\endcsname{\color[rgb]{1,0.3,0}}%
      \expandafter\def\csname LT8\endcsname{\color[rgb]{0.5,0.5,0.5}}%
    \else
      \def\colorrgb#1{\color{black}}%
      \def\colorgray#1{\color[gray]{#1}}%
      \expandafter\def\csname LTw\endcsname{\color{white}}%
      \expandafter\def\csname LTb\endcsname{\color{black}}%
      \expandafter\def\csname LTa\endcsname{\color{black}}%
      \expandafter\def\csname LT0\endcsname{\color{black}}%
      \expandafter\def\csname LT1\endcsname{\color{black}}%
      \expandafter\def\csname LT2\endcsname{\color{black}}%
      \expandafter\def\csname LT3\endcsname{\color{black}}%
      \expandafter\def\csname LT4\endcsname{\color{black}}%
      \expandafter\def\csname LT5\endcsname{\color{black}}%
      \expandafter\def\csname LT6\endcsname{\color{black}}%
      \expandafter\def\csname LT7\endcsname{\color{black}}%
      \expandafter\def\csname LT8\endcsname{\color{black}}%
    \fi
  \fi
    \setlength{\unitlength}{0.0500bp}%
    \ifx\gptboxheight\undefined%
      \newlength{\gptboxheight}%
      \newlength{\gptboxwidth}%
      \newsavebox{\gptboxtext}%
    \fi%
    \setlength{\fboxrule}{0.5pt}%
    \setlength{\fboxsep}{1pt}%
\begin{picture}(5102.00,3400.00)%
    \gplgaddtomacro\gplbacktext{%
      \csname LTb\endcsname
      \put(814,704){\makebox(0,0)[r]{\strut{}$0$}}%
      \csname LTb\endcsname
      \put(814,1134){\makebox(0,0)[r]{\strut{}$40$}}%
      \csname LTb\endcsname
      \put(814,1565){\makebox(0,0)[r]{\strut{}$80$}}%
      \csname LTb\endcsname
      \put(814,1995){\makebox(0,0)[r]{\strut{}$120$}}%
      \csname LTb\endcsname
      \put(814,2426){\makebox(0,0)[r]{\strut{}$160$}}%
      \csname LTb\endcsname
      \put(814,2856){\makebox(0,0)[r]{\strut{}$200$}}%
      \csname LTb\endcsname
      \put(946,484){\makebox(0,0){\strut{}$0$}}%
      \csname LTb\endcsname
      \put(1573,484){\makebox(0,0){\strut{}$2$}}%
      \csname LTb\endcsname
      \put(2199,484){\makebox(0,0){\strut{}$4$}}%
      \csname LTb\endcsname
      \put(2826,484){\makebox(0,0){\strut{}$6$}}%
      \csname LTb\endcsname
      \put(3452,484){\makebox(0,0){\strut{}$8$}}%
      \csname LTb\endcsname
      \put(4079,484){\makebox(0,0){\strut{}$10$}}%
      \csname LTb\endcsname
      \put(4705,484){\makebox(0,0){\strut{}$12$}}%
    }%
    \gplgaddtomacro\gplfronttext{%
      \csname LTb\endcsname
      \put(209,1941){\rotatebox{-270}{\makebox(0,0){\strut{}radius, $\mu$m}}}%
      \put(2825,154){\makebox(0,0){\strut{}time, $\mu$s}}%
      \csname LTb\endcsname
      \put(3718,1317){\makebox(0,0)[r]{\strut{}We = 336}}%
      \csname LTb\endcsname
      \put(3718,1097){\makebox(0,0)[r]{\strut{}RALEF 0.5 g/cm$^3$}}%
      \csname LTb\endcsname
      \put(3718,877){\makebox(0,0)[r]{\strut{}2D (uniform density)}}%
    }%
    \gplbacktext
    \put(0,0){\includegraphics{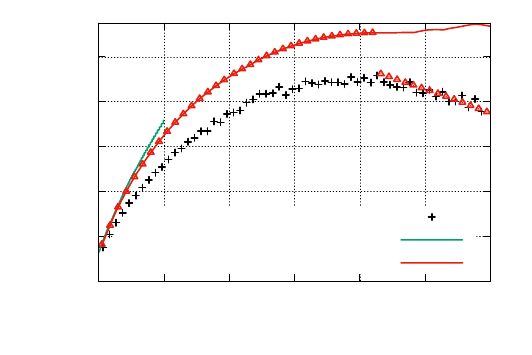}}%
    \gplfronttext
  \end{picture}%
\endgroup
    \caption{Target radius evolution over time. ARCNL We336 experiment (black dots) and calculations (color curves).}
    \label{Radii_We336}
\end{figure}

\begin{figure}[!htb]
    \centering
    \includegraphics[width=0.9\linewidth]{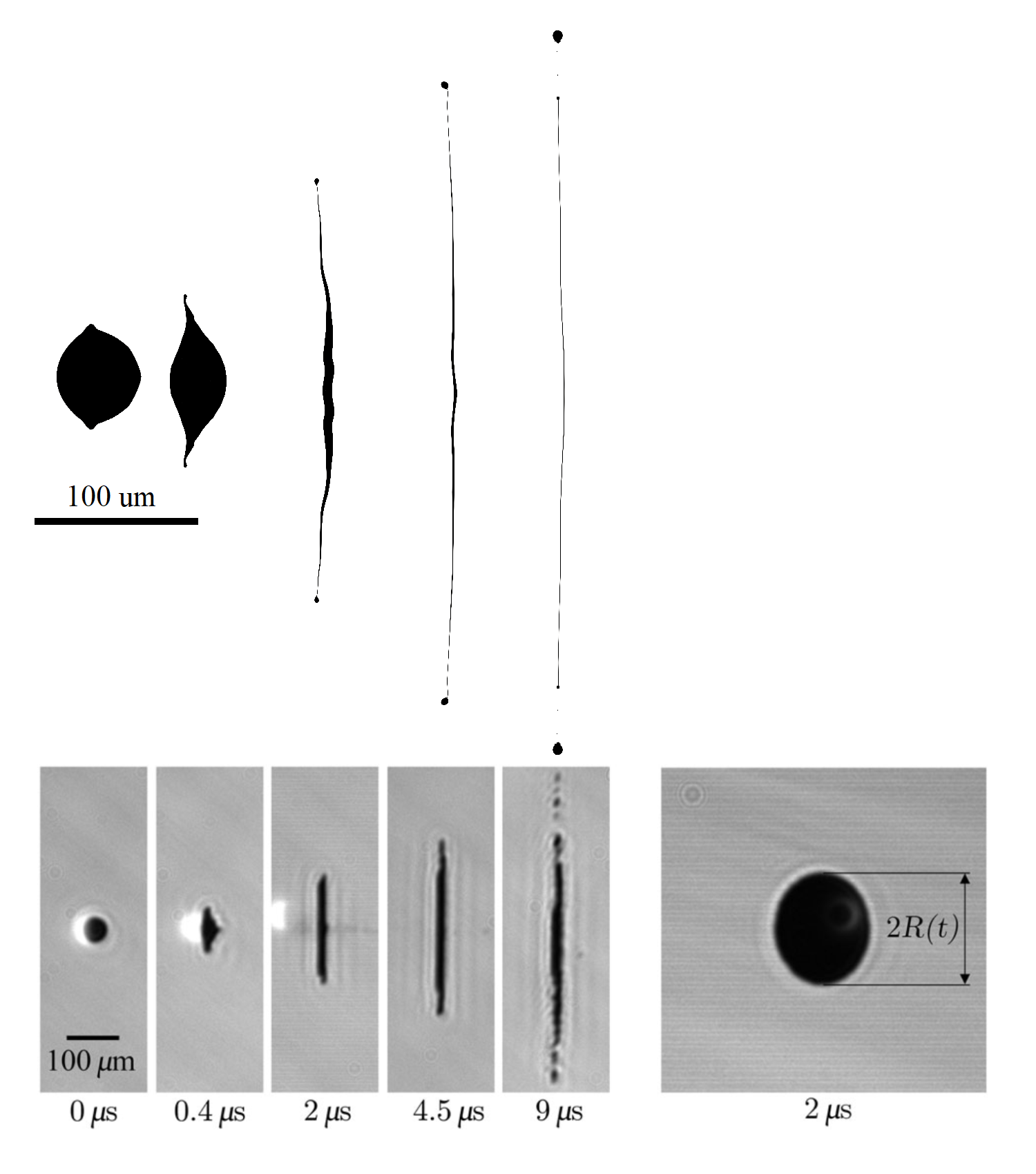}
    \caption{Target deformation after exposure by a laser pulse (ARCNL experiment We336).}
    \label{of_We336}
\end{figure}

\section{Summary}
A numerical study of the liquid tin drop deformation process into a disk after exposure by a laser pulse with an average intensity of about 10$^9$ W/cm$^2$ was carried out. For this purpose, a special modeling technique was developed and implemented using joint sequential use of RALEF-2D and OpenFOAM codes.

A complex model of radiation gas dynamics, the interaction of laser radiation with matter and thermal conductivity allowed us to adequately describe the first stage of EUV generation, including the absorption of laser radiation by a liquid target first and then the plasma formed as a result of laser ablation, the appearance of a weak shock wave propagating inside the target, formation of a field of velocities and densities, leading to deformation of the droplet into a disk-shaped target.

After the laser pulse is turned off, after several tens of nanoseconds, the contribution of the processes of radiation exchange and thermal conductivity becomes negligible. In this case, the surface tension forces begin to influence the deformation of the inertia-moving target.

The calculations showed good agreement with the experiment conducted at the ARCNL center at the University of Amsterdam (Netherlands).

\section*{Acknowledgment}
Calculations have been performed at HPC MVS-10P (JSCC RAS) and HPC K100 (KIAM RAS).
We thank Mikhail Basko for proofreading the manuscript and for fruitful discussions.


\bibliography{elsarticle-RALEF-OF}

\end{document}